\documentclass[sigconf]{acmart}

\AtBeginDocument{%
  \providecommand\BibTeX{{%
    \normalfont B\kern-0.5em{\scshape i\kern-0.25em b}\kern-0.8em\TeX}}}

\setcopyright{acmcopyright}

\usepackage{booktabs}
\usepackage{subfigure}

\definecolor{dark-green}{rgb}{0,0.6,0}

\definecolor{amaranth}{rgb}{0.9, 0.17, 0.31}
\definecolor{bostonuniversityred}{rgb}{0.8, 0.0, 0.0}
\definecolor{brightpink}{rgb}{1.0, 0.0, 0.5}
\definecolor{darklava}{rgb}{0.28, 0.24, 0.2}
\definecolor{darkgreen}{rgb}{0.0, 0.2, 0.13}
\definecolor{coolblack}{rgb}{0.0, 0.18, 0.39}
\definecolor{blue-violet}{rgb}{0.54, 0.17, 0.89}

\copyrightyear{2022}
\acmYear{2022}
\setcopyright{acmlicensed}\acmConference[IUI '22]{27th International Conference on Intelligent User Interfaces}{March 22--25, 2022}{Helsinki, Finland}
\acmBooktitle{27th International Conference on Intelligent User Interfaces (IUI '22), March 22--25, 2022, Helsinki, Finland}
\acmPrice{15.00}
\acmDOI{10.1145/3490099.3511152}
\acmISBN{978-1-4503-9144-3/22/03}

\newcommand{\gt}{{\color{coolblack}{\textit{GreaseTerminator}}}}

\begin{document}

\title[Mind-proofing Your Phone with GreaseTerminator]{Mind-proofing Your Phone: \\ Navigating the Digital Minefield with GreaseTerminator}



\author{Siddhartha Datta}
\email{siddhartha.datta@cs.ox.ac.uk}
\affiliation{%
  \institution{University of Oxford}
  \country{United Kingdom}
}

\author{Konrad Kollnig}
\email{konrad.kollnig@cs.ox.ac.uk}
\affiliation{%
  \institution{University of Oxford}
  \country{United Kingdom}
}

\author{Nigel Shadbolt}
\email{nigel.shadbolt@cs.ox.ac.uk}
\affiliation{%
  \institution{University of Oxford}
  \country{United Kingdom}
}

\renewcommand{\shortauthors}{Datta, Kollnig, and Shadbolt}

\begin{abstract}
Digital harms are widespread in the mobile ecosystem. As these devices gain ever more prominence in our daily lives, so too increases the potential for malicious attacks against individuals.
The last line of defense against a range of digital harms~--~including digital distraction, political polarisation through hate speech, and children being exposed to damaging material~--~is the user interface.
This work introduces \gt{} to enable researchers to develop, deploy, and test interventions against these harms with end-users.
We demonstrate the ease of intervention development and deployment, as well as the broad range of harms potentially covered with \gt{} in five in-depth case studies.
\end{abstract}


\maketitle

\section{Introduction}
\label{intro}





Mobile devices are becoming an ever larger invader of our digital wellbeing worldwide. The average time spent per day on mobile devices was 122 minutes in 2018, compared to only 39 minutes on desktop; usage time has doubled since 2014~\citep{stat_0}. 
As many as 95\% apps from certain categories on the Google Play Store were found to contain \textit{dark patterns} (malicious interface design to nudge users into performing actions they would not originally intend to do); 75\% of surveyed users could not detect dark patterns in their apps with certainty~\citep{10.1145/3313831.3376600}. 

The methods that malicious actors (e.g. app developers, social media bots, advertisers, and other end-users) can adopt to
influence user behavior can be seen as an extension of Cybenko et al.'s concept of \textit{cognitive hacking}~\citep{Cybenko_2002}.
They defined this as a computer or information system attack that `modifies certain user behaviors in a way that violates the integrity of the entire user information system'.
Examples of digital harms faced by end-users include
the gamification of apps (digital addiction),
the engineering of choice architecture to nudge users towards certain in-app decisions (dark patterns), and
the propagation of carefully-crafted information can incite problematic beliefs and emotions on social network platforms (disinformation, hate speech, self-harm).

While there exists a range of digital harms intervention tools, most of these tools are aimed at desktop browsers, such as ad-blockers 
and user-scripting tools. 
Few tools exist for mobile devices.
One reason for this absence of intervention tools is the relative absence of techniques for researchers to develop, deploy and test new interventions against digital harms with end-users.
This is why our work, rather than developing new interventions, proposes a framework, called \gt{}, for the development, assessment and deployment of new interventions.
This shall ultimately help researchers to
understand better how digital harms affect end-users, and what interventions against these harms are effective.

\noindent{\small \textbf{Contribution: }}
This works contributes \gt{} as an adaptable and robust app modification framework for researchers to analyse how a wide range of digital harms may affect end-users, and what countermeasures might be effective. 
Our current architecture provisions a set of \textit{hooks}, that use state-of-the-art machine learning (ML) methods, and can be used as ready-to-use templates to develop interventions against digital harms easily.
\gt{} enables researchers in machine learning, security, and human-computer interaction to test, deploy, and scale their theoretical models and machine learning architectures in the wild and put their algorithms into the hands of end-users, so as to study how to mitigate digital harms.
We demonstrate the improvements to intervention development, as well as the broad range of harms potentially covered with \gt{} in five in-depth case studies.


\noindent{\small \textbf{Structure: }}
The rest of this paper is structured as follows. 
In Section \ref{motivation}, we motivate our work, focusing on interface-based harms in smartphones.
In Section~\ref{rw}, we provide an overview of the current landscape of interventions, and derive requirements to overcome the limitations of the status quo. In Section~\ref{gt}, we explain the architecture of \gt{}. In Section~\ref{method}, we introduce our methodology to evaluate our implementation against our requirements. In Section~\ref{eval}, we evaluate \gt{} with a set of case studies of five possible interventions, and turn to our discussion and limitations in Section~\ref{dis}.
We
share
our conclusions in Section~\ref{conclusions}.
\section{Motivation: Digital harms in smartphones}
\label{motivation}

Before analyzing how to improve upon existing app modification frameworks, we motivate our work by briefly describing the variations of digital harms in smartphones.
The spectrum of harms and dangers that exist in the digital world is broad and highly individual. Different terms have been used interchangeably throughout literature. This is partly because different harms may be perpetrated by different actors (e.g. app developers, app users, state-owned bots), have different attack objectives (e.g. increasing revenue, rigging elections, inciting self-harm), and whether an attack requires malicious intent (e.g. 
users may unknowingly harm other users when sharing certain content).
While it is beyond this work to characterise all possible digital harms (instead, our work ultimately aims to help other researchers to identify, analyse and mitigate digital harms), two categories of harms are particularly relevant:

\textit{Developer-originated harms}
are harms crafted by developers of mobile apps. Based on the desired biases and actions of users that the app developer would like to influence, the developer can craft a flow of user interactions with interface elements. 
The objective of the developer tends to be revenue-oriented (e.g. maximizing usage time of the app, clicking ads, making in-app purchases). 
Many of these harmful design patterns can be summarized under Gray et al.'s 5-class taxonomy of dark patterns~\citep{10.1145/3173574.3174108}, including
\textit{interface interference} (elements that manipulate the user interface to induce certain actions over other actions), 
\textit{nagging} (elements that interrupt the user’s current task with out-of-focus tasks)
\textit{forced action} (elements that introduce sub-tasks forcefully before permitting a user to complete their desired task), 
\textit{obstruction} (elements that introduce subtasks with the intention of dissuading a user from performing an operation in the desired mode),
and \textit{sneaking} (elements that conceal or delay information relevant to the user in performing a task).

\textit{User-originated harms}
are executed by participants of a community-enabled app. Examples include 
users and advertisers/trolls on a social networking app,
buyers and sellers on e-commerce sites, or
content creators and content viewers on media sharing platforms.
Harms can appear as content payloads, such as text, image or video, and they may or may not be created with the intention of harming other users.
Online harms perpetuated by other users may include 
heavily-biased content (e.g. disinformation, hate speech),
self-harm (e.g. eating disorders, self-cutting, suicide),
cyber crime (e.g. cyber-bullying, harassment,
promotion of and recruitment for extreme causes (e.g. terrorist organizations),
demographic-specific exploitation (e.g. child-inappropriate content, social engineering attacks)~\citep{hmgov, 10.1145/2998181.2998224, 10.1145/3038912.3052555, 10.1145/3313831.3376370, 10.1145/3359186}.


This brief overview of digital harms in smartphones underlines that the user interface serves as a last line of defense against many of the potential mobile harms.
This is why, in this work, we are interested in harms that transpire through the user interface, and how these can be analysed by academic researchers who are interested in developing effective interventions against these harms.

\section{Background: App modification frameworks}
\label{rw}

\subsection{Existing digital harms interventions}

Since there are many mobile-based harms, there, too, exist many existing frameworks to tackle harms in mobile apps.
The scope of our following review includes both end-user and research tools. We aim to understand the strengths and weaknesses across these tools and derive key requirements for a harms intervention system.
These existing frameworks fall into six different categories.




\noindent
\textbf{{\color{bostonuniversityred}{\textit{(1) App-external modifications}}}}~\citep{kovacs_thesis,bodyguard,10.1007/s10916-013-0001-1,10.1145/2675133.2675244, 10.1145/2968219.2971591,10.1145/2858036.2858403,10.1145/2541831.2541870,al,10.1145/3229434.3229463} are the arguably most common way to change default app behaviour.
An end-user would install an app so as to affect other apps.
No change to the operating system or the targeted apps is made. An uninstall of the app providing the modification would revert the device to the original state.
This approach is widely used to track duration of app usage, send notifications to the user during usage (e.g. timers, warnings), block certain actions on the user device, and other aspects.
The {\color{bostonuniversityred}{\textit{HabitLab}}}~\citep{kovacs_thesis} is a prominent example developed by Kovacs et al. at Stanford.
This modification framework is community-driven and open-source, and provides separate interventions for desktop and mobile devices.

\noindent
\textbf{{\color{amaranth}{\textit{(2) App pre-install modifications}}}}~\citep{gd,apkm,jeon_dr_2012,rasthofer_droidforce_2014,davis_retroskeleton_2013,garcia-alfaro_appguard_2014,Aurasium,lp,Davis12i-arm-droid:a} make changes to app source code before installation on the user's device.
A prominent example is {\color{amaranth}{\textit{AppGuard}}}~\citep{garcia-alfaro_appguard_2014}, a research project by Backes et al. that allowed users to improve the privacy properties of apps on their phone.
A popular solution in the community is the app {\color{amaranth}{\textit{Lucky Patcher}}}~\citep{lp} that allows to get paid apps for free, by removing the code relating to payment functionality directly from the app code.
Another example is {\color{amaranth}{\textit{GreaseDroid}}}~\citep{gd} that allows to remove dark patterns from apps through a community-driven app modification approach.
    
\noindent
\textbf{{\color{darkgreen}{\textit{(3) OS-level modifications}}}}~\citep{cydia,xp,agarwal_protectmyprivacy_2013,enck_taintdroid_2010,InstaPrefs} make use of the highest level of privilege escalation to make modifications to the operating system and apps as a root user.
On iOS, {\color{darkgreen}{\textit{Cydia Substrate}}}~\cite{cydia} is the foundation for jailbreaking and further device modification. A similar program, called {\color{darkgreen}{\textit{Xposed Framework}}}~\citep{xp}, exists for Android.
Other prominent examples are {\color{darkgreen}{\textit{ProtectMyPrivacy}}}~\citep{agarwal_protectmyprivacy_2013} for iOS and {\color{darkgreen}{\textit{TaintDroid}}}~\citep{enck_taintdroid_2010} for Android, which both extend the functionality of the smartphone operating system with new functionality for the analysis of apps' privacy features.
    
\noindent
\textbf{{\color{blue-violet}{\textit{(4) Virtual environment modifications}}}}~\citep{vx} create a virtual environment on the mobile device with simulated privilege escalation. Users install apps into this virtual environment and apply tools of other modification approaches that may require root access (e.g. app run-time modifications). 
An example is {\color{blue-violet}{\textit{VirtualXposed}}}~\citep{vx}.
    
\noindent
\textbf{{\color{darklava}{\textit{(5) Browser modifications}}}}~\citep{oilcan,tm,swipe} make change to client-side site code when a site is rendered on mobile web browsers. This approach tends to port tools that exist on desktop browsers.
Examples include
{\color{darklava}{\textit{OilCan}}}
~\citep{oilcan},
{\color{darklava}{\textit{TamperMonkey}}}
~\citep{tm} and
{\color{darklava}{\textit{Swipe for Facebook}}}
~\citep{swipe}.
    
\noindent
\textbf{{\color{coolblack}{\textit{(6) Visual overlay modifications}}}} render graphics on an overlay layer over any active interface, including browsers, apps, videos, or any other interface in the operating system. The modifications are visual, and do not change the functionality of the target interface. 
It may render sub-interfaces, labels, or other graphics on top of the foreground app.
Examples are \gt{}, {\color{coolblack}{DetoxDroid}}~\citep{detoxdroid},
{\color{coolblack}{Gray-Switch}}~\citep{grayswitch} and {\color{coolblack}{Google Accessibility Suite}}~\citep{accessibility}.




\subsection{Evaluating existing app modification approaches}


We now evaluate limitations of existing app modification frameworks to identify a gap in interventions development.
There are at least two important barriers to implementing interventions on mobile devices: 
(i) the difficulty for the developer to develop the intervention and reach users (e.g. developing individual interventions, scaling each intervention implementation across a breadth of apps and platforms), and 
(ii) the difficulty for the user to adopt the intervention (e.g. initial set-up cost for each intervention, managing all the different interventions and interoperability).


For interventions that make use of 
{\color{amaranth}{\textit{app pre-install modifications}}}, or {\color{darkgreen}{\textit{OS-level modifications}}}, some of the interventions may require the end-user to gain admin/root access on their mobile device (and in some cases void warranty) in order to use the intervention. Some of these tools may not even be available on the Google Play Store.
 For intervention developers, a much higher level of expertise is required to develop these interventions, since they interfere with the low-level components of the Android system or Android apps.
{\color{blue-violet}{\textit{Virtual environment modification}}} methods avoid the privilege escalation harm for the aforementioned modification frameworks; however, it requires the user to execute all their mobile device activity within this virtual environment, such as using browsers or apps. 
It also poses the same difficulties for intervention developers.
For interventions that rely on the {\color{darklava}{\textit{browser modification}}} framework,
these have lower barriers to development as they build upon established development languages, foremost JavaScript.
Such browser-based interventions are easy to use for end-users, but suffer from not being deployable to all in-app harms.

The issue of difficult implementation of interventions on mobile is further illustrated by {\color{bostonuniversityred}{\textit{HabitLab}}}~\citep{kovacs_thesis}, 
which allows expert users to construct interventions that apply to all sites or to specific sites (e.g. scroll-locks, time usage limitations). 
Despite efforts in patching digital harms on the browser platform, there is a mismatch in offerings available as of
date of writing this manuscript
despite a community of expert users actively contributing to the project source code.
The current mismatch in offerings and range of considerations in development and user adoption indicates the uphill battle that digital self-control developers face with respect to portability, that patches constructed on one platform cannot easily be ported to another platform. 
The additional consideration that different mobile operating systems require different modification frameworks to deploy interventions adds an extra layer of complexity to interventions design and deployment.


Once an intervention developer has selected an app modification framework, they will be challenged by the volume of interventions required to implement. 
As of 2020, the Google Play Store provisions more than 2.9 million Android apps~\citep{playstore}. 
Each app varies by version; developers update apps and source code.
{\color{amaranth}{\textit{GreaseDroid}}} supports both app-agnostic (apply to all apps, regardless of the app or its version) as well as app-specific (only work for a specific app) interventions. 
{\color{amaranth}{\textit{Lucky Patcher}}} performs app modifications on-device, and caters to users interested in getting access to paid apps for free.
For interventions to be app-agnostic, the interventions developer needs to find patterns through static code analysis for many apps to automate the process of modifying code.
In addition, mobile-based operating systems are \textit{multi-interfaced}. This means there are multiple interfaces where digital harms can reside within a single device, ranging from text/image/video within an interface, mobile browsers, apps, and the operating system as a whole.
Interventions that directly remove interface elements may obstruct access to certain app functionality. For example, removing the newsfeed entirely so that a user cannot see hate speech or targeted ads can also mean they cannot catch-up on life updates of friends~\citep{erad}. Though interventions may be needed to mitigate user-originated harms, we may not need to block all user-generated content on Facebook such as those that are non-malicious.
In the case of {\color{bostonuniversityred}{\textit{HabitLab}}}~\citep{10.1145/3274364, kovacs2021now}, users uninstalled or weakened the extent of interventions over time, possibly due to the intrusive nature of the tool to accessing desired functionality. 
Some self-control apps focus on providing diagnostics (e.g. informing the user of the problem during usage rather than blocking the interface elements), and relying on the user to take action. Lyngs et al.~\citep{lyngs_self-control_2019} presented an indication that if the interventions felt too directive or controlling, then people may wish to implicitly rebel against the interventions. 



\subsection{Requirements for an intervention assessment framework}

From the above review, we identify two \textit{requirements} that our framework, aimed at the assessment of new and existing interventions, should fulfill to address important gaps in the existing literature, and to analyse a broad range of digital harms:

\begin{itemize}
    \item {\small \textbf{(Requirement 1) Interface-orientation}}: The app modification framework should intervene against different interfaces in a multi-interfaced system (including media, mobile browsers, apps, and operating system).
    \item {\small \textbf{(Requirement 2) Ease of Use}}: 
    The app modification framework should lower barriers to usage and development for users and developers respectively. 
    For end-users, this could mean not requiring expert skills. 
    For developers, this could mean simplifying the process of implementing an intervention and of scaling interventions across other similar interfaces.
\end{itemize}

\noindent
We implement these requirements in our framework that we discuss in Section \ref{gt}. 
The various intervention solutions using the different app modification frameworks have been developed with different requirements, and the requirements we list here do not indicate the aforementioned solutions are sub-optimal, but merely that there is a gap that we would like to fill with \gt{}.
\section{GreaseTerminator}
\label{gt}

\subsection{System architecture}


\begin{figure*}[]
    \centering
    \includegraphics[width=0.8\linewidth]{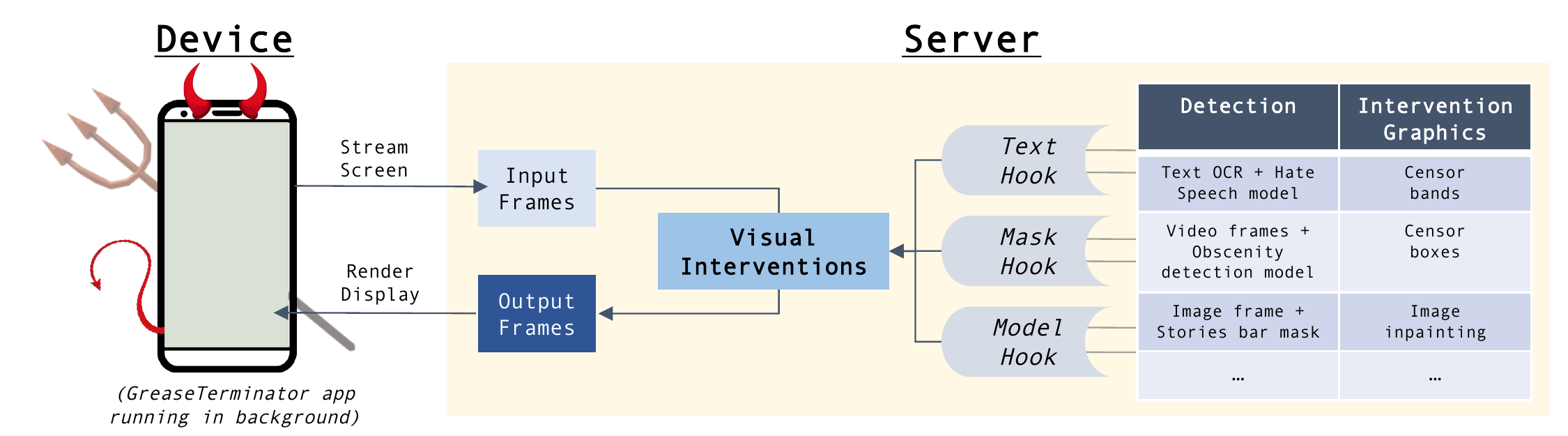}
    \caption{Architecture of \gt{}.}
    \label{fig:archi}
\end{figure*}

\noindent
The concept of \gt{} is to render interventions in the form of overlay graphics based on detected elements. Rather than implementing program code changes natively, we make use of Android's overlay capabilities to draw a translucent overlay over the user's display, and on this overlay layer we render graphics meant to intervene with content and elements detected on the user's screen. 
The interventions are generalizable in that they are not app-oriented, but \textit{interface-oriented}. 
Interventions do not target specific apps, but general interface elements and patterns that could appear across different interface environments. 
The overlays can be drawn over many interfaces, including
apps, mobile browsers, media content, and on the operating system as a whole.
Within the \gt{} ecosystem, we denote three stakeholders: \textit{app developers} who created the app to intervene with, \textit{intervention developers} who create interventions to be deployed through \gt{}, and \textit{end-users} who use apps and need to use interventions to protect themselves from digital harms.
Since we envisage that \gt{} will be used as a tool for researchers to assess different interventions, intervention developers will usually be researchers.

To illustrate an application of the framework, we provide a sample prototype containing an Android app, scripts to run the server, and modules for all the interventions evaluated\footnote{Source code: \url{https://github.com/greaseuniverse/greaseterminator}}.
The current implementation (Figure~\ref{fig:archi}) requires server-side computation to evaluate display contents and compute the overlay. The user first installs the \gt{} app on their Android device. The app streams the user's device screen to an external server through the Android Debug Bridge (ADB), a debugging tool provided by Google.
On the external server, there exist a set of interventions pre-crafted by intervention developers that rely on a set of \textit{hooks}. These interventions parse the streamed image for its content (e.g. graphical elements, text), and dynamically render interface components
(e.g. filter masks, warning text). The output of each intervention would be a render of its corresponding graphics onto the translucent overlay (after the user has granted overlay permissions at the start), which would then be streamed back onto the user's device.

The key advantages of a server-based approach are that we can use existing ML tools for PCs, and leverage the usually high processing power of PCs as compared to many smartphones.
While a server-based approach might suffer latency issues (more in Section~\ref{lat}) when used by users outside their homes,
the approach is suitable for lab studies, and when users are at home and can install and run the server architecture on their own computers (or alternatively, have a low-latency internet connection and can connect to a central server).




\subsection{Hooks: Enabling Intervention Development}
\label{3.2}

To simplify intervention development for researchers,
we provide a range of pre-built \textit{hooks}, which serve as ready-to-use templates for intervention development.
Hooks can be used in conjunction with other hooks, or be embedded as components in the logic of intervention scripts.

The developer (i.e. a researcher, or a software engineer on a research team) first installs a \gt{} server instance. The \gt{} app installed on the end-user's smartphone would then stream the user UI to this server. The server would, in turn, stream a modified interface back. 
Developers can craft scripts on how to process the incoming stream of images on the server using hooks: these both perform pattern detection within and modification to images. 
\gt{} provides an easy-to-use default set of pre-developed text/mask/model hooks.
Alike to module/library imports in Python,
developers call functions defined by a hook,
provide input arguments (e.g. model weights)
and return outputs on to a user or another hook.
It is possible to implement a range of further hooks. However, we found that these three support a wide range of interventions.

\begin{figure}[ht]
    \centering
    \includegraphics[width=\linewidth]{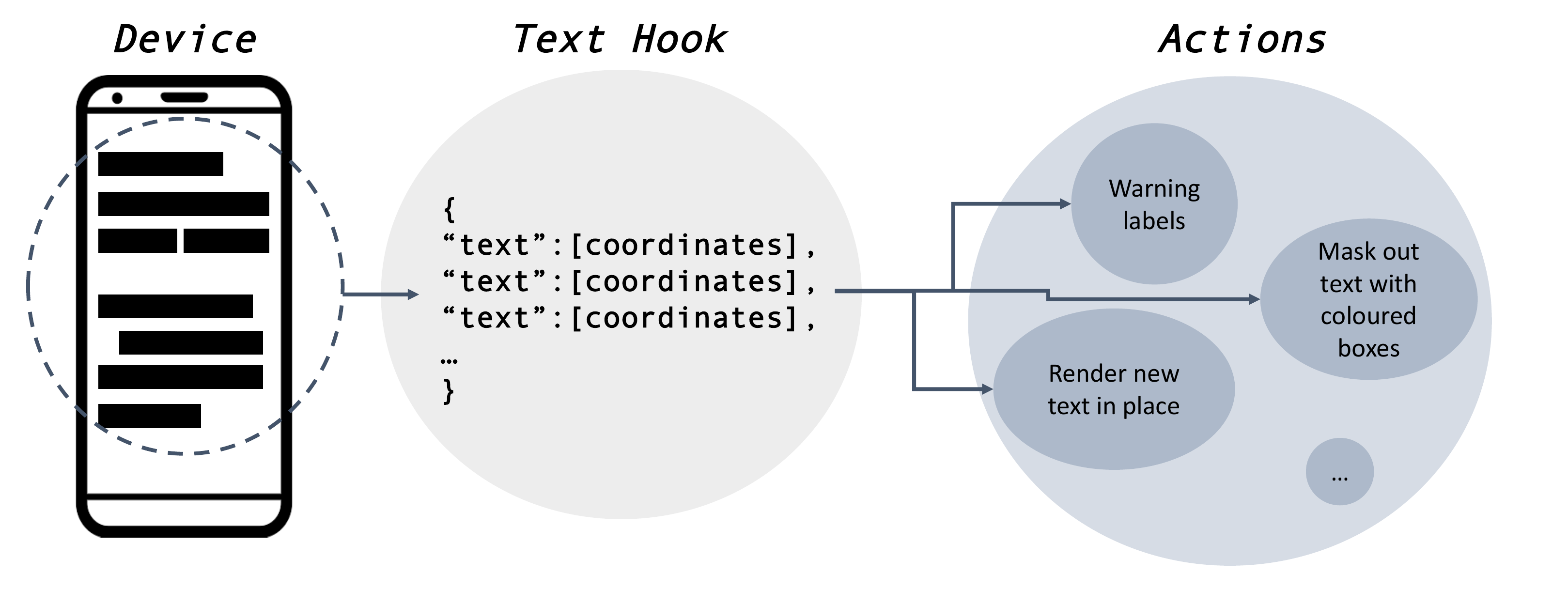}
    \caption{\textit{[Text]} Detection \& processing of on-screen text.}
    \label{fig:hook1}
\end{figure}

\subsubsection{Text Hook}
The text hook enables modifying the text that is displayed on the user's device (Figure~\ref{fig:hook1}).
We implement this through character-level optical character recognition (OCR) that takes the screen image as an input and returns a set of characters and their corresponding geometric coordinates. 
We first identify regions of interest with EAST text detection~\citep{zhou2017east}. The model detects text in images and returns a set of regions with text. Then we use Tesseract~\citep{tesseract} to extract characters within each region containing text.
With the availability of real-time textual data from each image instance, an intervention developer can store information processed by the image to be processed by subsequent models via the model hook. For example, one could construct a model that identifies information being used cross-app (e.g. targeting of ads on Facebook based on searches executed on Google Chrome). Another example is the intervention developer may perform real-time text filtering that renders black filter boxes on text with certain conditions (e.g. hate speech, disinformation). 

\begin{figure}[ht]
    \centering
    \includegraphics[width=\linewidth]{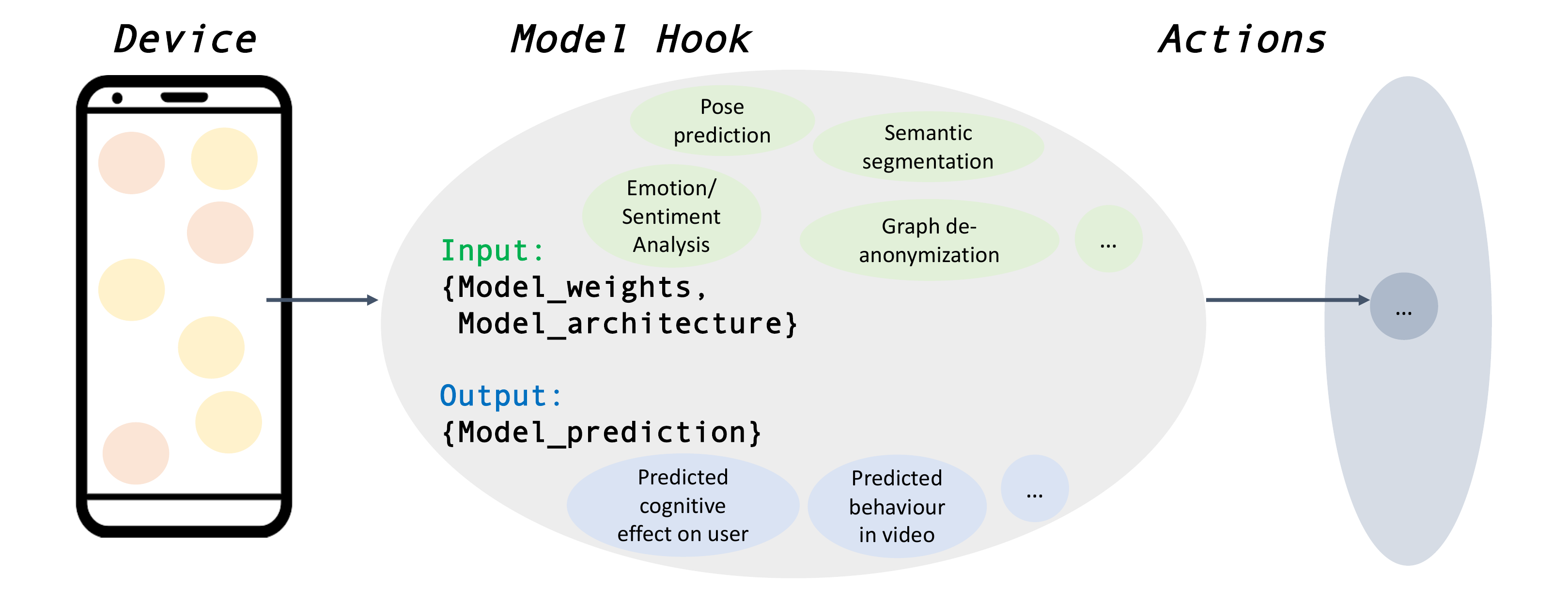}
    \caption{\textit{[Model]} Versatile model inference \& deployment.}
    \label{fig:hook3}
\end{figure}

\subsubsection{Model Hook}
A model hook loads any machine learning model to take any input and generate any output (Figure~\ref{fig:hook3}).
This allows an intervention developer to embed models (i.e. model weights and architectures)
to inform further overlay rendering.
We can connect models trained on specific tasks (e.g. person pose detection, emotion/sentiment analysis) to return output given the screen image (e.g. bounding box coordinates to filter). This output can then be passed to a defined rendering function (e.g. draw filtering box).

\begin{figure}[ht]
    \centering
    \includegraphics[width=\linewidth]{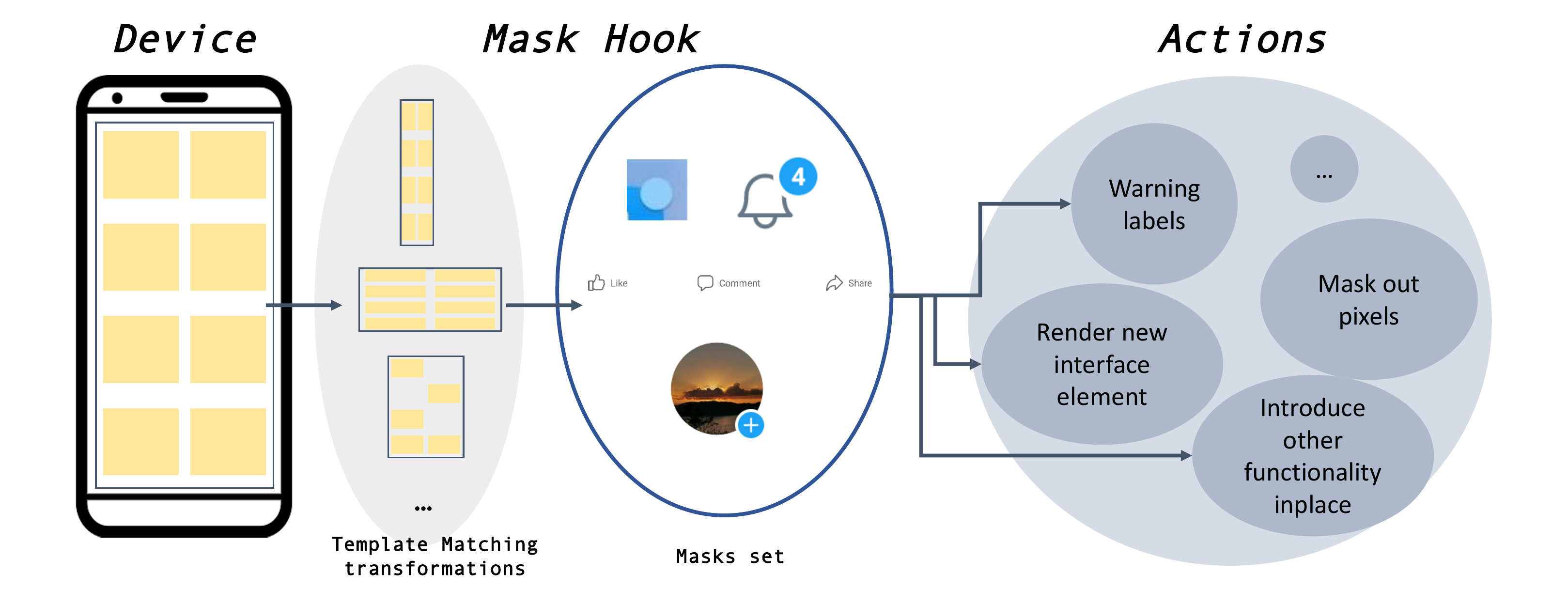}
    \caption{\textit{[Mask]} Detection \& processing of GUI elements. }
    \label{fig:hook2}
\end{figure}

\subsubsection{Mask Hook}
The mask hook matches the shown screen content against a target template of multiple images,
such as cropped screenshots of problematic content (Figure~\ref{fig:hook2}).
To do this,
we implement \textit{multi-scale multi-template} matching by resizing an image multiple times and sampling different subimages to compare against each instance of mask in a masks directory (where each mask is a cropped screenshot of an interface element). 
Masks can be collected by developers (or crowd-sourced with risk of backdoor attacks \citep{datta2021widen, datta2022backdoors}).
Sometimes the masks can operate adaptably between apps; for example, if we create a mask of the notification badge in the corner of an app on the homescreen \includegraphics[height=0.25cm]{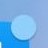}, we can identify similar badges in the corners of other apps, and thus remove notification badges at an operating system level.
The mask hook could be connected to rendering functions such as highlighting the interface element with warning labels, or image inpainting (fill in the removed element pixels with newly generated pixels from the background).
Developers can tweak how the hook is applied, for example using the matching algorithm with contourized images (shapes, colour-independent) or coloured images depending on whether the mask contains (dynamic) sub-elements, or using few-shot learning models if similar interface elements are non-uniform.


\subsection{Latency}
\label{lat}

An important challenge with our \gt{} prototype is latency.
Our current implementation of \gt{} computes interventions off-device to reduce the time for evaluating the different hooks, given the larger processing power of computers over smartphones.
This rests on the assumption that many useful research studies can be done in lab settings, or when users are at home and can run the server functionality from their own computers.

Latency arises in two forms: 
\textit{model execution latency} (e.g. arising from the size of model, type of input, number of models needed, computing on server or device, computing on GPU or CPU), and
\textit{network latency} to send and receive data from the device (e.g. arising from the protocol used, network speed and stability, size of data packets).
Throughout this paper, we measure this latency as the time difference between when the new rendered screen image is shown minus when the original input screen was shown.

To reduce model latency, we use a fast implementation for mask hooks, providing two state-of-the-art image inpainting techniques: (i) majority-pixel inpainting, and (ii) Fast Marching image inpainting~\citep{journals/jgtools/Telea04}. 
The latter method along with deep learning based image inpainting algorithms have large architectures with higher processing load per image instance compared to the former, and suitable for non-uniform images with varying features. The delay may accumulate, so instead we recommend intervention developers to opt for the default majority-pixel inpainting method, where it finds the most common colour value in a screen image, and inpaints with that colour rather than generating a set of pixels. 
As many mobile interfaces are standardized or uniform or consistent from a design perspective compared to images from the natural world, this may work in many instances.
To enable mask detection with a single instance for mask hooks, while there have been advances in one-shot learning (such as Siamese Networks~\citep{Koch2015SiameseNN}, Model-Agnostic Meta Learning \citep{finn2017modelagnostic}, or Learn2Weight \citep{datta2021learnweight}), these networks require high training and inference times. If trained with low frequency, this method may be feasible; however, for our mask hook functionality we need a more scalable method to quickly perform one-shot detection (to detect mask coordinates on screen) for image inpainting with new mask images, thus we implement multi-scale multi-template matching. Once adjusted for contours and colour-scaling, the uniformity of app interfaces allows for more accurate template matching than if applied to the natural world.
With more computational bandwidth (i.e. more models with negligible latency), we can use other work to augment \gt{} functionality: UIED~\citep{10.1145/3368089.3417940} is an example of recent work that uses CNNs to detect GUI elements on mobile interfaces, and could be used to improve element detection and inpainting to complement multi-scale multi-template matching;
%
%
UIAutomator is another example where, though it cannot extract metadata from sub-interfaces such as videos or images, it can extract metadata of UI elements such as text and UI hierarchy, and this can complement the provision of input data.

To reduce network latency, we also tried to use the Android ScreenCast functionality that allow devices to send images to external servers over TCP/UDP.
However, we found that using ADB (50 milliseconds) to perform screen relay has the significantly lower latency compared to TCP/UDP screen relays with ScreenCast functionality (800 milliseconds).
We also tested on-device model inference (i.e. zero network latency), running a Linux machine on an Android device and execute deep learning models in Python on this installed machine so as to avoid the redevelopment time needed 
to rewrite models from Python into JAVA/Kotlin.
Unfortunately, the model latency is roughly 60--120 seconds per inference, which is not feasible for use in practice.
However, there are methods to distil and simplify model architectures to perform inference on mobile devices faster and with less computational load~\citep{44873}, but we leave this to future work.

With \gt{} being a new research framework for other follow-up research, 
the assessment of specific ML models and their impact on latency is somewhat beyond the scope of this paper.
However, we present a methodology for measuring and evaluating latency for future work.
%
Visible latency should be of concern when the framerate of the underlying screen content increases (e.g. game or video), or the number of required models and size per model increases.
In our current setup, we used a GPU-enabled server (4x NVIDIA GeForce RTX 2080), relayed images through an ADB connection (download/updated speed $\sim 250$Mbps, with each image sent and received by the server amounting to $\sim 1$Mb), and optimized the used models wherever possible (e.g. tested alternative algorithms and suitable parameters).
The theoretical latency per one-way transmission should be 
$\frac{1 \times 1024 \times 8 \textnormal{bits}}{250 \times 10^6 \textnormal{bits/s}}$ = 0.033ms.
Assuming a server instance would use only one of our GPUs and with reference to existing online benchmarks \citep{aib}, the latency for 1 image (CNN) and text (LSTM) model would be 5.1ms and 4.8ms respectively. The total theoretical latency for our setup should thus be ($2 \times 0.033 + 5$) = 5.07ms. This supports a maximum framerate of 197fps, sufficient to cover the framerate of videos (24--60fps), and sufficient to cover the framerate of normal device use (e.g. reading, scrolling).
Depending on the use case, the intervention developer can optimize the hooks scripts further; for example, during normal use a large number of models can be run (40 models if the user is browsing at 5fps), and 8 models can be run for 25fps video.

Overall, in our current implementation, our device updated the screen image for deployment of the mask roughly every 50 milliseconds, which is hardly noticeable at run-time.
However, as the complexity of the used models increases, latency might increase, so models must be used with care.

\vspace{-0.25cm}
\section{Method}
\label{method}


\gt{} is evaluated against its requirements (introduced in Section~\ref{rw}): interface-orientation and ease of use. To do this, we need to first determine a set of test interventions from previous work or literature. Then, we need to observe the changes to the intervention development process. These findings are illustrated in Section~\ref{eval} through four metrics: {\small \textbf{Problem}}, {\small \textbf{Known implementations}}, {\small \textbf{Challenges}}, and {\small \textbf{Implementation with \gt{}}}. Then, we evaluate the effect of \gt{} on the framework requirements and on different stakeholders in the intervention development process in Section~\ref{dis}.

We first select a set of interventions that we would like to implement with \gt{}. These interventions are chosen such that they represent a wide range of digital harms that can be found online, considering different types of digital harms, origins of the harm (e.g. developers, users), mediums in which the harm may manifest (e.g. text, video, app).
As part of this, we illustrate the \textit{problem} that an intervention tries to solve, and existing \textit{implementations} of achieving this purpose.
We also raise any empirical evidence of user studies demonstrating the effectiveness of such interventions if applicable.
The criteria and interventions selected are shown in Table~\ref{fig:res}.

Having selected these interventions, we then turn to the question of how these can be implemented in \gt{}, and what benefits this brings.
We consider the cognitive process and explicit actions taken by the intervention designer and developer to deploy interventions on a multi-interfaced operating system.
Given a known intervention to a known set of digital harms, we acknowledge the \textit{challenges} that an intervention developer would have in implementing an intervention designer's proven intervention without \gt{}, and show the reduced barrier to development and deployment in an \textit{implementation} with \gt{}.
The evaluation of the development process with respect to development cognition and development actions are based on work in cognitive walk-throughs~\citep{10.1145/223904.223962, 10.1145/223355.223735} and software engineering ethnographic studies~\citep{10.1145/3338906.3338976}.

It should be made clear that the contribution of this work is not the novelty or performance of intervention designs.
\gt{} assists in helping developers assess intervention ideas, but the performance of the intervention may be dependent on non-framework factors.
Some models may not be contextually-accurate; false positives/negatives with models indicate needed improvement with the model, but does not indicate a failed intervention design or app modification framework.
Some models may be very large (e.g. transformers) and increase latency (which can be combated by distilling models into smaller models, e.g. network pruning/distillation \citep{44873}), while some models may have lower accuracy than others in test-time.
If researchers use \gt{}, they should design studies such that participants are specifically evaluating interventions based on their design and requirements, and not based on deployment limitations.
End-user acceptability on this shift in interaction is recommended further study.
Additionally, the selection of interventions that we use to evaluate \gt{} may suffer from different biases. 
Some evaluated interventions may also be focused on showing the breadth of interventions development with \gt{}, but we do not conclude from the selection of interventions that this specific intervention is best implemented through \gt{}. 
For example, performing image matching to detect scrolling to implement usage time intervention (Section \ref{5.2}) could be replaced with an alternative implementation with input method extension (IIRC).
To demonstrate the viability of the tool to deploy models quickly, we would be biased towards available models/architectures rather than developing novel deep learning architectures.
By focusing on interventions already discussed in the literature, this research introduces \gt{} as a novel research methodology to develop, deploy, and assess interventions for end-users.

\begin{table*}
\small
\begin{tabular}{p{.18\textwidth}p{.06\textwidth}p{0.16\textwidth}p{.1\textwidth}p{.15\textwidth}p{.15\textwidth}p{0.25\textwidth}}
	\toprule
	\textbf{Intervention} & \textbf{Hook} & \textbf{Tackled Harm} & \textbf{Origin} & \textbf{Interfaces} & \textbf{On mobile?} \\ \midrule
	Element occlusion (stories bar) & Mask & Distraction & Developer, User & Browser, App & \citep{swipe}, \citep{InstaPrefs}, \citep{gd} \\ 
	Demetrification & Mask & Distress & Developer & Browser, App & No, but desktop: \citep{fbd}, \citep{twd}, \citep{igd} \\ 
	Hate Speech Filter & Text & Polarisation & User & Text, Image & \citep{bodyguard} \\ 
	Media Moderation & Model & Kids' Wellbeing & User & Image, Video & No \\ 
	Usage locking & Model & Distraction & Developer & Browser, App, Operating System & \citep{10.1007/s10916-013-0001-1}, \citep{10.1145/2675133.2675244}, \citep{10.1145/2968219.2971591}, \citep{10.1145/2858036.2858403}, \citep{10.1145/2541831.2541870}, \citep{al}, \citep{10.1145/3229434.3229463} \\ \bottomrule
\end{tabular}
\caption{Interventions already implemented with \gt{}, aiming to tackle a broad variety of harms.}
\label{fig:res}
\end{table*}

\section{Interventions against Digital Harms}
\label{int}
\label{eval}

Digital harms are highly individual, and it is difficult to design a robust strategy to cope with these harms, as discussed in Section~\ref{rw}.
We now introduce a set interventions that 1) have been examined in the academic literature, 2) are widely used in practice, and 3) cover a range of potential harms.
Table~\ref{fig:res} shows the interventions selected and implemented.
We provide the pseudocode for interventions; our prototype implementation of server deployment and interventions are in Python (though other languages/setup should suffice), and the screenshots appended in figures are non-cherry-picked screenshots of the implemented interventions running in real-time.

\subsection{Element Occlusion}


For the phenomenon of digital distraction, we focus on two commonly used interventions:
element occlusion from social media, and usage locking.
Social media apps feature many elements designed to increase usage time, and thereby advertising revenue, which these two interventions try to overcome.
We first focus on element occlusion, and then turn to usage locking in the next subsection.




\noindent{\small \textbf{Problem: }}
Many interface elements (such as the \textit{stories bar} on Instagram and LinkedIn, the news articles on Snapchat, or the \textit{find friends} option on Facebook)
are \textit{obstruction} dark patterns~\citep{10.1145/3173574.3174108}, introduced as a sub-task during the user experience to induce a user to enter a continuous loop of consuming content and lengthening the immersion time in the app.
These dark patterns introduce addictive behavior of remaining on the app longer than intended, and priming users to enter such loops each time the user enters the app.
Sometimes, as in the case of stories bar on Instagram, they enter full-screen to redirect the user from their initial task with this new sub-task of watching videos, and shift attention away from their original intended usage of the app.


\noindent{\small \textbf{Known implementations: }}
There is a limited supply of tools that assist users with the removal of elements in mobile apps.
The app {\color{darklava}{\textit{Swipe for Facebook}}}~\citep{swipe} can remove the stories bar on Facebook, {\color{darkgreen}{\textit{Instaprefs}}}~\citep{InstaPrefs} can disable the auto-advance feature and disable ads in Instagram stories, and {\color{amaranth}{\textit{GreaseDroid}}}~\citep{gd} can remove the stories bar for apps with crafted patches (e.g. Twitter).
There is indication that the removal of the entire feed interface (which contains the subtasks) does reduce digital distraction
\citep{lyngs_i_2020}.

\noindent{\small \textbf{Challenges: }}
An example of an element that exists across many Android apps is the \textit{stories bar}. 
Though there are efforts for each individual app, there is no generalizable solution to patch the stories bar in all apps.
Removal of this element for each app requires careful inspection of each app's source code and making modifications through trial and error. 
Even if an intervention developer identifies the specific code segments pertaining to the stories bar in a specific app, it may not be similar to the stories bar source code segment in another app, even with similar design, or after app updates, thus the effort needs to be repeated many times.

\begin{figure*}
    \centering
    \subfigure[Implementation pseudocode]{
    \includegraphics[height=2in]{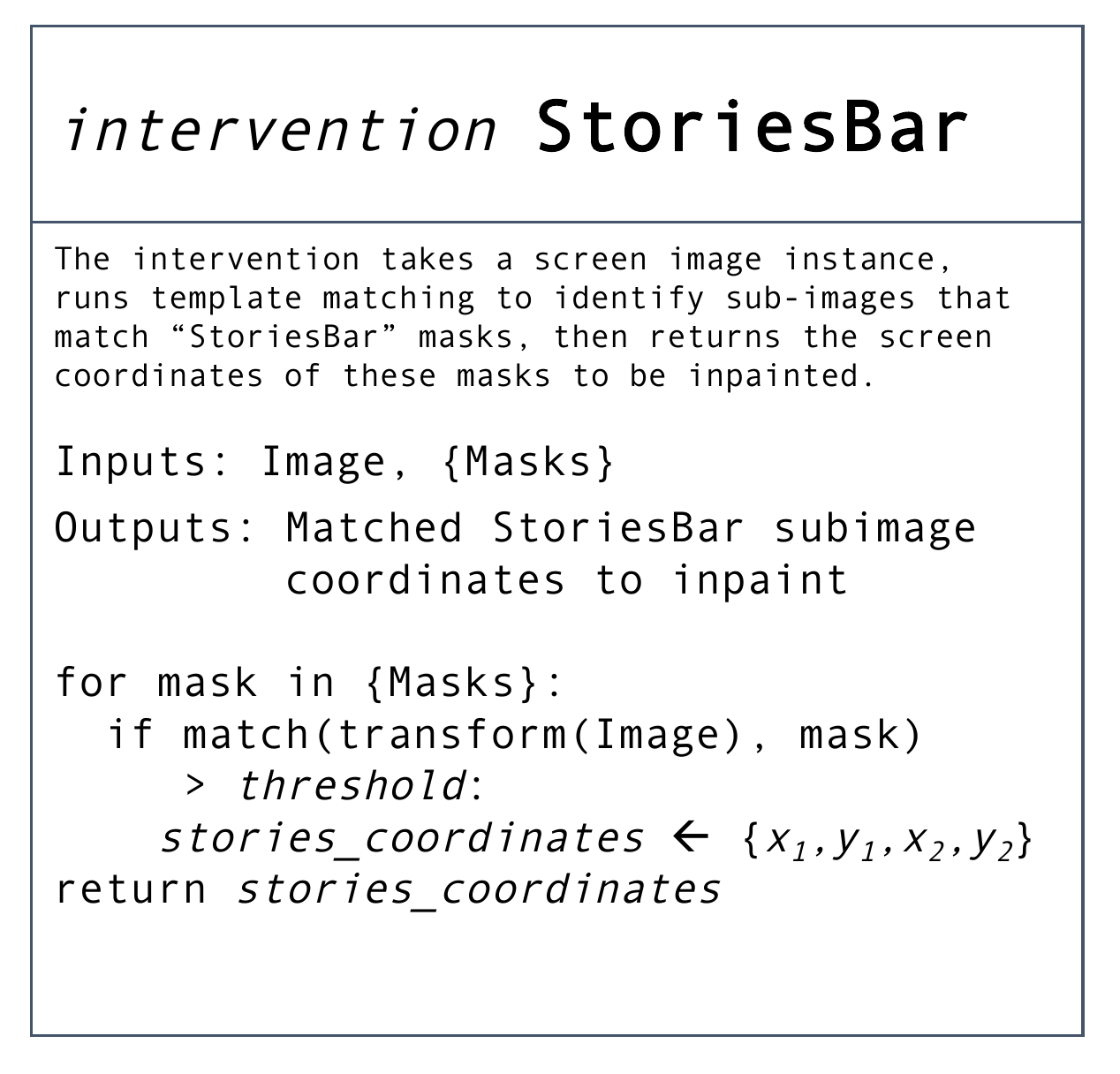}
    }\hfill
    \subfigure[Twitter (before \& after)]{
    \includegraphics[height=2in]{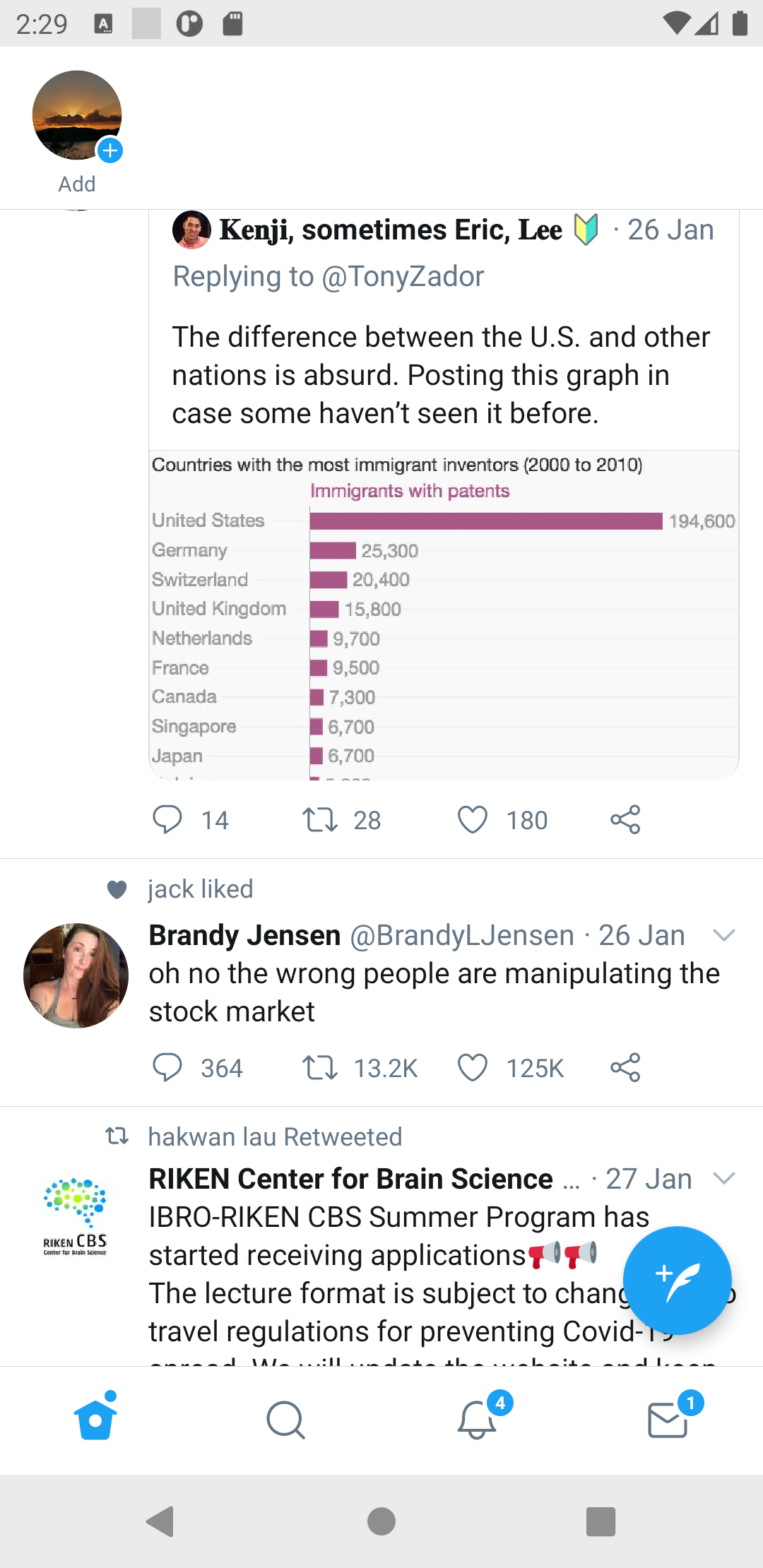}
    \includegraphics[height=2in]{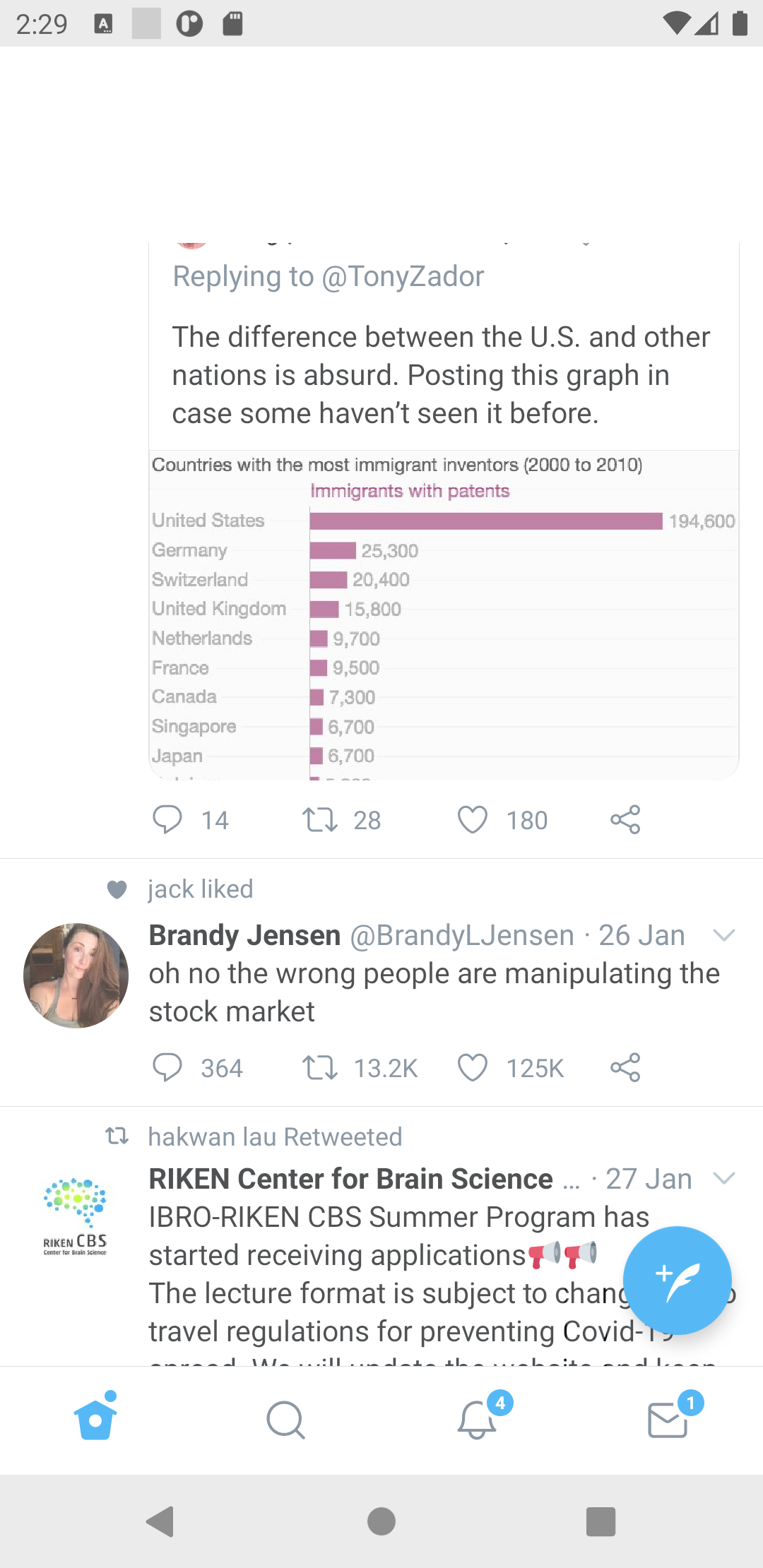}
    }\hfill
    \subfigure[LinkedIn (before \& after)]{
    \includegraphics[height=2in]{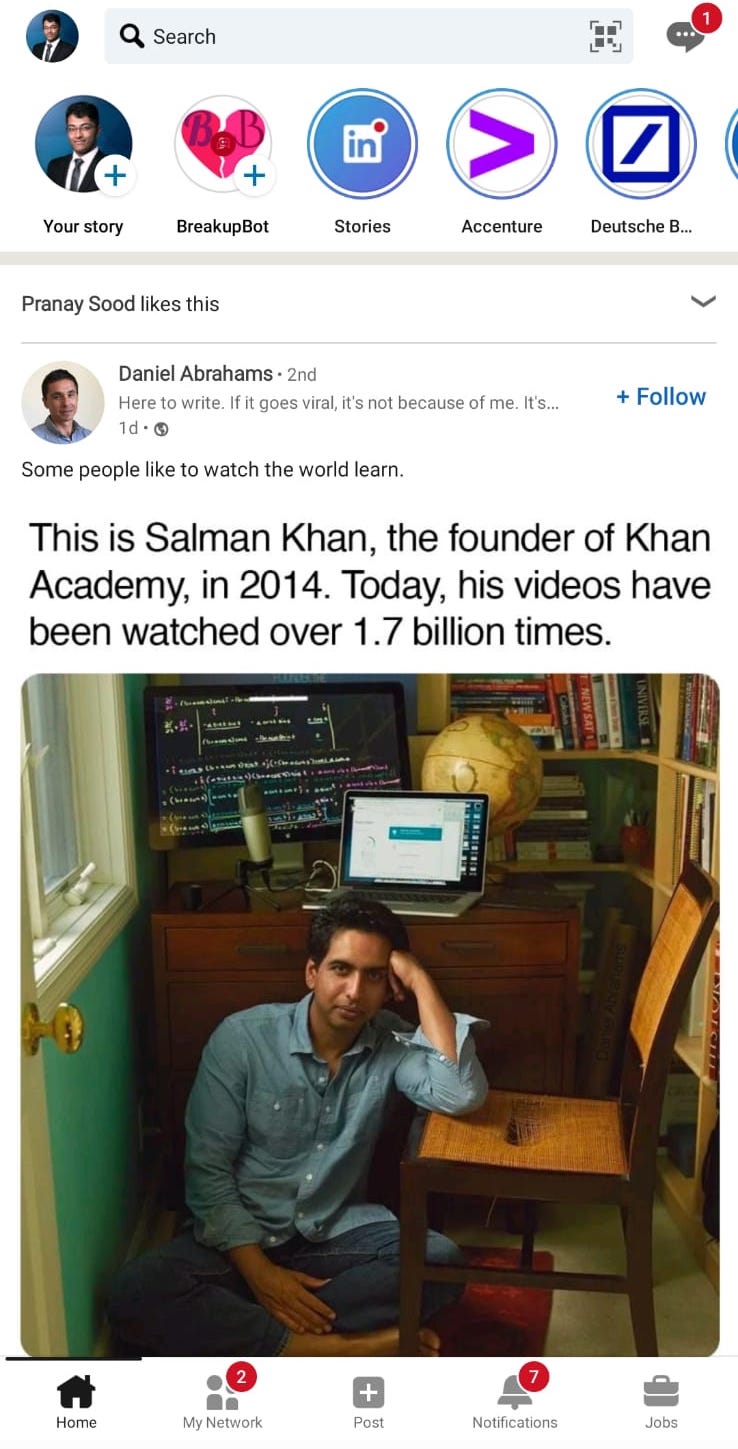}
    \includegraphics[height=2in]{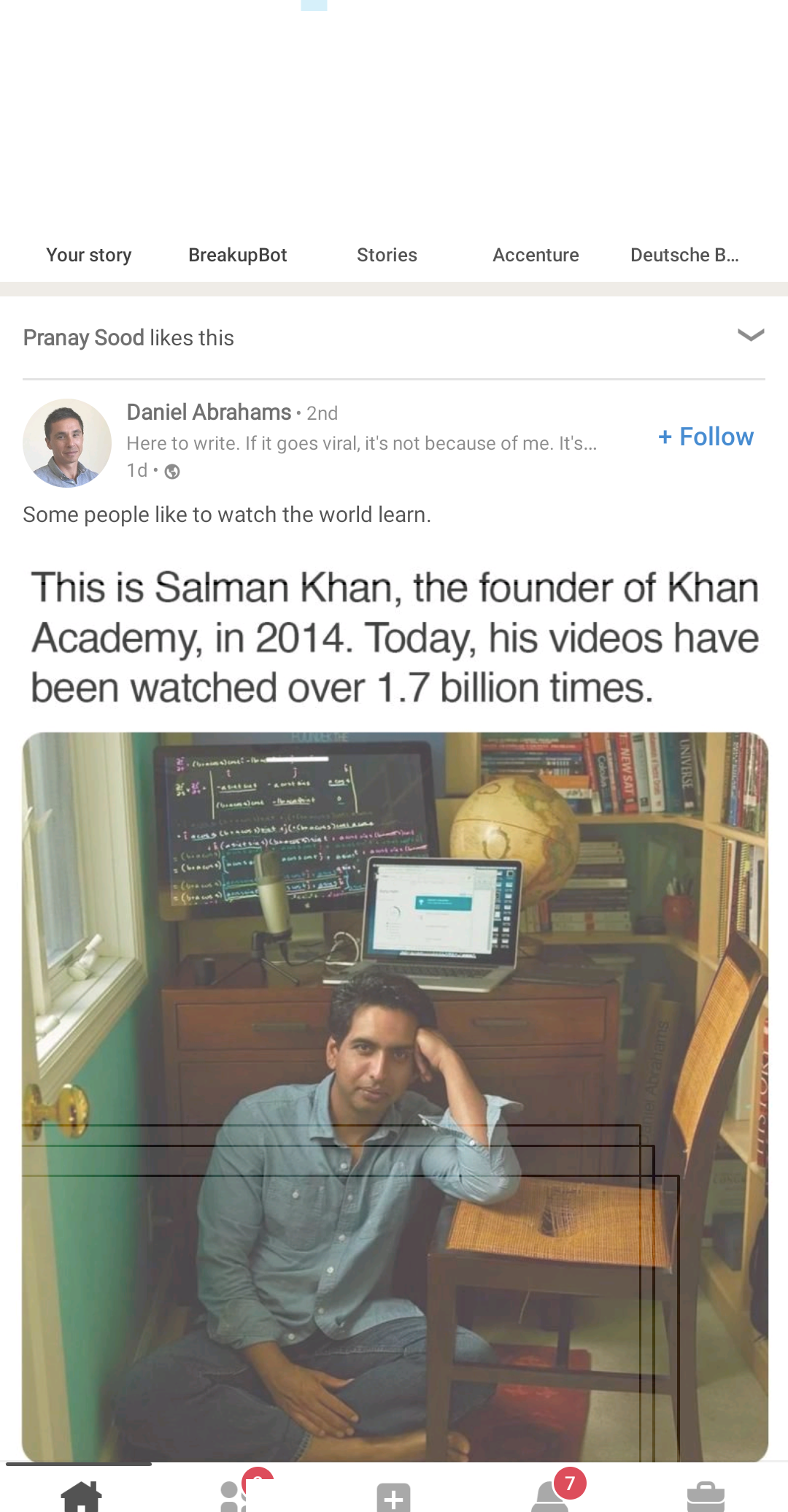}
    }
    \caption{Removal of the stories bar with \gt{}, thereby potentially increasing the feeling of control over one's app use and reducing time spent.}
    \label{fig:storiesbar}
\end{figure*}

\noindent{\small \textbf{Implementation with \gt{}: }}
All an intervention developer must do is use the mask hook of \gt{}, and provide a screenshot of the stories bar to remove the stories bar from an app.
In this case, as the content inside the circles of the stories bar is dynamic, we would need to compare the contours (edges, shapes) of the stories bar mask (a set of aligned circles) against the contours of the user's screen. 
The mask hook would compute the likelihood of regions in the user screen matching the contours of the stories bar mask provided.
Contours can be automatically extracted from the stories bar image; here we used Suzuki's contour tracing algorithm \citep{SUZUKI198532}, accessible through the opencv library.
This then allows to reduce its visibility (Figure~\ref{fig:storiesbar}), or remove it altogether.
Such an implementation may generalise across app updates, the browser-based version of an app, and different apps with stories bar functionality, as this contour pattern (a set of aligned circles) is common.
Even if this is not the case, it is simple to provide additional screenshots for these cases.
If app developers make modifications to the stories bar interface, intervention developers can quickly respond by taking a screenshot of the new stories bar, and add to the masks database. 

\subsection{Usage Locking}
\label{5.2}




The second digital distraction intervention we consider is usage locking, the gradual `locking' of the user interface from usability, which has been implemented across a range of self-control apps.

\noindent{\small \textbf{Problem: }}
The longer an end-user is using an app, the longer they are exposed to the other digital harms that persist within the app. Some interfaces curate content in an \textit{infinite-scroll} form, and users keep scrolling for extended periods of time.
This gives rise to digital addiction as well as facilitating other digital harms.

\noindent{\small \textbf{Known implementations: }}
There are two types of interventions tackling the issue of usage control.
The least intrusive
of the two is a time usage dashboard.
The intention of the tool is to show users how much time they have spent on each app, to warn users of addictive behavior, and relying on them to change usage patterns.
Examples of such interventions on Android that track usage and allows users to set usage goals include {\color{bostonuniversityred}{\textit{SAMS}}}~\citep{10.1007/s10916-013-0001-1}, {\color{bostonuniversityred}{\textit{NUGU}}}~\citep{10.1145/2675133.2675244}, {\color{bostonuniversityred}{\textit{Menthal}}}~\citep{10.1145/2968219.2971591}, {\color{bostonuniversityred}{\textit{MyTime}}}~\citep{10.1145/2858036.2858403}. 
The other usage control method is to reduce usability of the app once certain conditions have been reached by enforcing different techniques. Conditions may include a time limit (e.g. set as a goal by the user) or a certain number of scrolls. 
{\color{bostonuniversityred}{\textit{AppDetox}}}~\citep{10.1145/2541831.2541870} prevents a user from accessing certain apps once certain conditions (e.g. time of day, number of launches) have been reached. {\color{bostonuniversityred}{\textit{Auto Logout}}}~\citep{al} force-closes the app when a certain time limit has elapsed. {\color{bostonuniversityred}{\textit{Okeke et al.}}}~\citep{10.1145/3229434.3229463} evaluated the use of excessive vibrations to hinder usability of an app once a time limit has expired. 
There also exist a set of desktop self-control tools that disincentize further site usage.
{\color{coolblack}{\textit{Anchor}}}~\citep{anchor} reduces visibility of the screen by rendering a mask of the blue ocean at varying opacity levels as the user scrolls deeper into a page. {\color{bostonuniversityred}{\textit{Timewaste Timer}}}~\citep{twt} uses financial punishment to hinder usability, charging a user funds if they exceed the time limit.
Lyngs et al.~\citep{lyngs_self-control_2019} showed that the introduction of such usage control tools can reduce the time spent, number of daily scrolls, and number of Like operations on the Facebook site on laptop.
Kovacs et al.~\citep{10.1145/3290605.3300560} show with {\color{bostonuniversityred}{\textit{HabitLab}}} that usage control features not only reduce time spent on the intended site but also reduces time spent on other sites.

\noindent{\small \textbf{Challenges: }}
Self-control apps  exist that can measure the time elapsed per app and warn users of excessive use.
{\color{bostonuniversityred}{\textit{Okeke et al.}}}~\citep{10.1145/3229434.3229463} uses 
vibrations that physically warn the user to stop using the app.
An intervention developer could design a timer method that automatically closes an app once a time limit has been reached; the intervention developer could use {\color{amaranth}{\textit{GreaseDroid}}}~\citep{gd} to create an app-agnostic intervention that inserts this time-locking functions to each app of varying versions. This could be inserted into apps that render media (e.g. videos) as well as mobile browser apps. 

\begin{figure*}
    \centering
    \includegraphics[width=\linewidth]{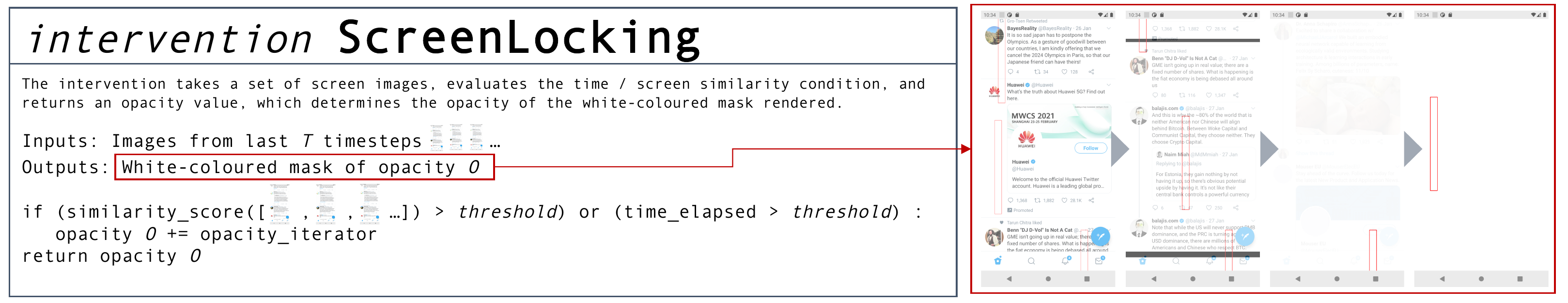}
    \caption{Usage Locking: Prevention of infinite scrolling with \gt{}. While there exists a range of existing tools that provide similar interventions, these usually do not aim to detect infinite scroll and take actions based upon this observation. As such, \gt{} promises varied new research directions.}
    \label{fig:lock}
\end{figure*}

\noindent{\small \textbf{Implementation with \gt{}: }}
An intervention developer can use the model hook of \gt{} to detect scrolling within an app, and increasingly reducing the 
visibility of the content of the app on the screen
once excessive scrolling (e.g. great number of scrolls or time of scrolling) is detected (Figure~\ref{fig:lock}).
For scrolling detection, we use a multi-scale multi-template matching model that segments different parts of the screen to compares against segments in images in the last \textit{T} timesteps, and additionally use average image hashing to compare segment similarity. 
If there is a shift in an identical element between two screen frames, then that would be defined as a `scroll'.
This approach can apply to any interface within the mobile device, such as time elapsed when watching a video, number of paragraphs/posts read in discussion forums on a mobile browser, number of stories viewed in social media apps, or even just scrolling through the homescreen.

\subsection{Demetrification}

\noindent{\small \textbf{Problem: }}
There is a great reliance on social media user interfaces indicating virality, including the number of likes, shares and comments. The metrification of created content helps social media platforms identify content that is likely to be of interest to their users either based on similarities in preferences in the users' peers, or simply by proportion of the population interested in it. 
Through the metrification of social media platforms, it is proposed that our patterns of interactions are altered~\citep{fbdemet} into wishing to optimize our social interactions based on these metrics, e.g. maximizing the number of friends, maximizing the number of likes our created content receives. This perpetuates how we perceive content posted online (e.g. news posted on Twitter with a high number of retweets), how we perceive friendships (e.g. which friends liked our photos), and how addicted we are to using an app (e.g. refreshing the likes count).
Consequently, it has been shown that the inclusion of such metrics on social media platforms induces anxiety and distress, particularly in adolescents~\citep{https://doi.org/10.1111/cdev.13422}.

\noindent{\small \textbf{Known implementations: }}
No demetrification tools exist on mobile devices at the moment.
There are a set of demetrification tools available on the desktop browser versions of social media platforms, namely the Facebook Demetrificator~\citep{fbd}, Twitter Demetrificator~\citep{twd}, and Instagram Demetrificator~\citep{igd}. These demetrification tools identify and remove the client-side source code pertaining to the numbers indicating the number of likes, shares and comments. 
It has been shown that the reduction of social media use resulted in subsequent reductions in levels of anxiety and depression~\citep{doi:10.1521/jscp.2018.37.10.751}.

\noindent{\small \textbf{Challenges: }}
Metric interface elements are specific to each app and would need to be customized for each app. 
They would be rendered repeatedly under each post, thus with some code analysis of the app code, it should be possible to implement a demetrification intervention for each social media app.
As each social media app have distinctly different metrics bar (to contain the like/share buttons), the intervention will need to be tailored for each app.
One concern is the duplication of efforts in demetrification required between the app version and mobile browser version of the social media platform.
An ease of use risk is the potential for app developers to obfuscate or change the code for the metrics such that an intervention for a specific version of the app does not work for another. 
Another ease of use risk is the possibility of the app developer performing server-side integrity checks for feed rendering or source code, and blocking functionality or user access if the check fails.
Though demetrification needs to be performed for each app individually, it should be robust to detection and defenses deployed by the app developer. 

\begin{figure*}
    \centering
    \includegraphics[width=\linewidth]{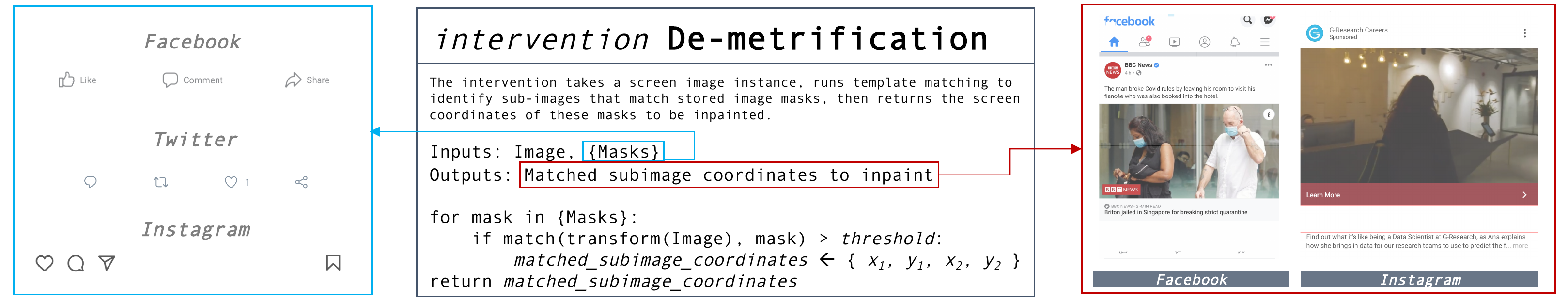}
    \caption{Demetrification of apps with single mask instances with \gt{}. While Instagram has publicly declared that it was planning to implement similar functionality into their main app, this has not yet been been deployed widely (if at all). \gt{} could help assess the reasons behind these decisions (e.g. that users spend less time on Instagram through this intervention, thereby seeing fewer ads and generating less revenue for Instagram, meanwhile increasing user happiness and satisfaction).}
    \label{fig:dm}
\end{figure*}

\noindent{\small \textbf{Implementation with \gt{}: }}
An intervention developer can easily remove metrification elements of an app, by using the mask hook of \gt{} and simply providing 
a cropped screen image of the metrics bar in this app (Figure~\ref{fig:dm}). We tested this for a range of social media apps (including Facebook, Twitter and Instagram).
Once a mask has been crafted for a specific app platform, the metric can potentially be removed for both the app and mobile browser version of the social media platform.
Though the intervention developer would still need to tackle each app individually, the time required for collecting screenshots for each app is very low.
If an intervention developer would like to avoid aggregating masks for each app, they could attempt at constructing a model that predicts the likelihood that a certain interface element is an app metric or not, and deploy it via the model hook.

\subsection{Hate Speech Filtering}
\label{sec:hatespeech}



An important segment of online harms is the propagation of inciting text content.
This can influence users based on multiple factors, such as based the frequency of exposure to a user, the number of friends who like or share or approve the content in the user's network, or even the relative demographic background of the user.
It is important to be able to filter the content or at least warn the user of the harmful nature of such content.
The social media companies tend to have a mandate that conflicts with this filter; they wish to maximize content creation and propagation but also minimize content that may have severe consequences (e.g. political fact-checking, hate speech), and this may create grey-area content spaces where the content is still likely to affect certain users (e.g. children) but are not filtered out by the platform. In these cases, interventions are needed to interfere unilaterally. 
Additionally, there exists a variety of disinformation and hate speech model architectures available for training and deployment.
Though some may argue against the efficacy of such automated detection models, part of the refinement of these models should be based on real-world deployment, which \gt{} could assist with. Thus, there is an indication of a gap between a problem and resources that could be matched.

\noindent{\small \textbf{Problem: }}
The target of hate speech tend to be either directed towards a specific person or entity, or generalized towards a group of people sharing a common protected characteristic~\citep{elsherief2018hate}, with visible linguistic differences (e.g. directed hate speech being more personal, informal, angrier). 
The way text content is presented can influence the biases and beliefs of a user, and the update in biases can be dependent on the styling of the text, the person who shared or promoted the text, the velocity at which the text is shared or liked, etc.
For example on the Gab network, hate speech on social network platforms tend to spread faster and farther to non-hate-speech-spreading-users, and tend to be generated by users of prominent positions on the network~\citep{10.1145/3292522.3326034, 10.1145/3415163}.
Given how dangerous text can be once it is released onto a network, it is important for it to be managed properly.

\noindent{\small \textbf{Known implementations: }}
Most efforts in hate speech intervention are partaken by the app platforms themselves. If app developers themselves propagate hate speech, the app store platforms such as Google Play Store have content moderation requirements~\citep{googledev} and filter these apps out. 
If the hate speech is user-generated, social network platforms have content moderation techniques deployed to either highlight text of disturbing/concerning nature or directly withdraw the content from the platform, e.g. Facebook~\citep{fbdev}, Twitter~\citep{twdev}. 
{\color{bostonuniversityred}{\textit{Bodyguard}}}~\citep{bodyguard} is a social media moderation app that streams comments and some user-generated content from social media apps into an external app, and filters out hateful speech.

\noindent{\small \textbf{Challenges: }}
Intervention developers can identify code segments that render text blocks, and add 
if-condition logic that chooses to not render text boxes if text inside the text box is considered as hate speech once processed on an external server through a trained hate speech model.
Content creators may attempt to bypass such interventions and circulate hate speech using images or video; interventions developers could first pass the media through optical character recognition to extract text, then render the media block if no hate speech is detected.
This approach would require the intervention developer to inspect each app individually, and separate intervention code would be required on mobile browsers. 
Another approach for an intervention developer would be to stream the content of social media apps into another external app, and performing the filtering on that external app, though this would limit app functionality, as the intervention developer would need to port every function from the original app as the version updates.

\begin{figure} 
    \centering
    \includegraphics[width=\linewidth]{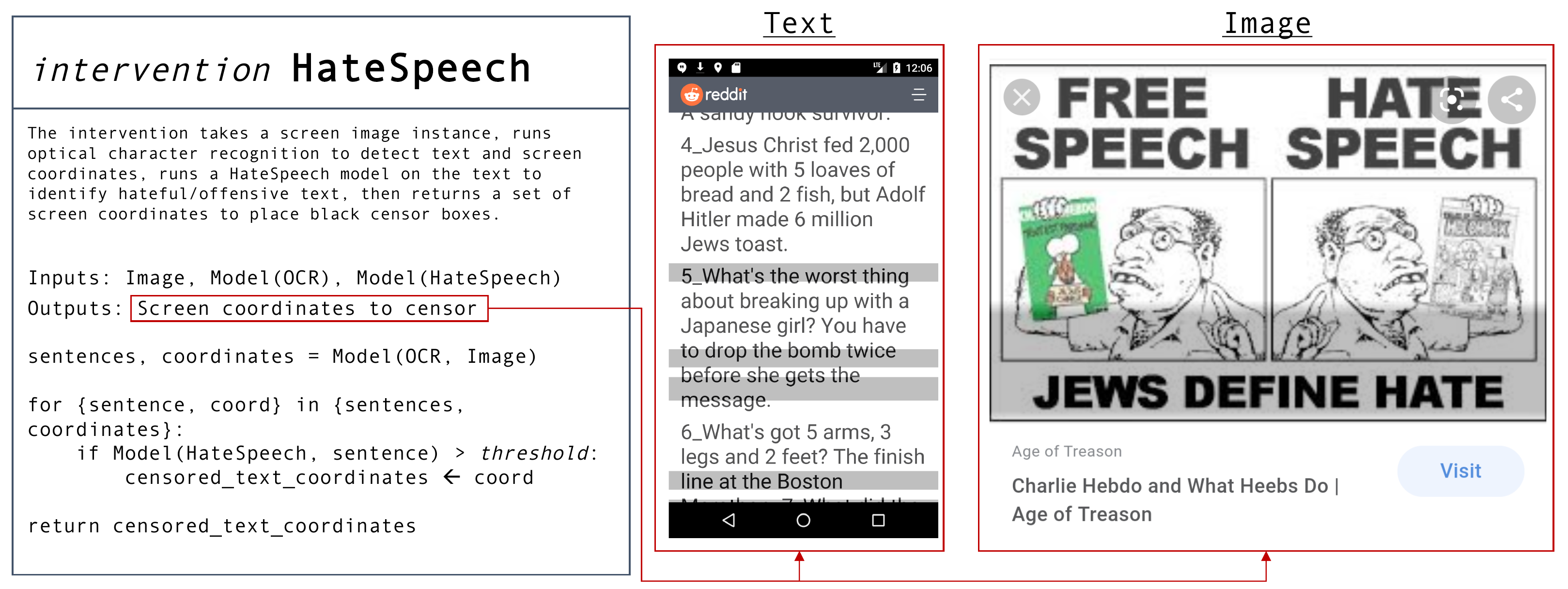}
    \caption{Hate Speech Detection \& Filtering with \gt{}. The used ML model clearly leads to \textit{overblocking}, but this can potentially be addressed through a more fine-grained model in the future. \gt{} aims to help researchers in finding better such models, with the help of end-users.}
    \label{fig:hs}
\end{figure}

\noindent{\small \textbf{Implementation with \gt{}: }}
To implement in-app hate speech detection and filtering (Figure~\ref{fig:hs}), the patch developer needs to use the text hook and choose an appropriate hate detection model.
We chose the hate speech model by Davidson et al.~\citep{hateoffensive}.
If the model evaluates some text to be hateful or offensive with a specific likelihood score and it exceeds the threshold set by the intervention developer, then they can assign an action to take place with respect to the screen coordinates of hate speech, such as filtering completely with black text boxes.
This method is robust to text found in browsers, inside apps (e.g. Facebook), outside apps (e.g. operating system homescreen), images (e.g. memes), and videos.
If we wish to extend hateful content filtering to other non-text forms such as image or video, we can make use of models hooks and implement image-to-text models~\citep{radford2021learning} to translate images into a set of text, and evaluate the text for the extent of hatefulness.
The development ease is high because the developed can rely on the provided text hook, and only needs to chose an appropriate hate model, of which there exist many.

\subsection{Media Moderation}

\noindent{\small \textbf{Problem: }}
Similar to heavily-biased text content, the exposure of visual content that is likely to update the biases or beliefs of an end-user may need moderation. Unregulated exposure of content that contain extreme views or biases may harm the user's actions.

For young children, there are filters on platforms to block age-inappropriate content. Content that may be harmful for viewers of varying ages pertain to nudity, use of weapons, blood and gore, extremist actions, etc. In Facebook’s 2019 community standards enforcement report, they mitigated 37.4 million pieces of content that violated their child nudity or child sexual exploitation policy~\citep{fbt}.
Unfortunately these filters may vary by platform depending on which country the app originated from or which region a user is situated, and the sensitivity of the filters of the platform. 

Though there are efforts by the app developers to control content flow, there is no universal standard, hence there is no guarantee that the efforts are uniform. Sometimes the user may wish to see the content of the video, and a small section of the video may be age-inappropriate, but platform owners may face liability in modifying video content created by a platform user.

Additionally, one of the biggest challenges in user-generated content is modifying or moderating videos in real-time. If a video is identified to be misaligned with a platform's moderation policies, then the social network platform will need to evaluate whether to delete the video. There may also be ways to bypass moderation depending on the algorithm used by the platform, such as limiting the indecent exposure to a few minutes out of a long video. Rather than relying on the ambiguity and uniformity across the content moderation policies of social network and video sharing platforms, unilateral action should be taken. 

\noindent{\small \textbf{Known implementations: }}
There are no known real-time video moderating or perturbing tools available to users. Video moderation is currently fully relied on the platforms. 

\noindent{\small \textbf{Challenges: }}
To modify video content against age inappropriate content, intervention developers could opt for either post-processed rendering or real-time rendering.
For the former, the intervention would first download the video, apply automated video editing operations, then render the new video for the user; as a new video is being processed and generated, latency would be quite high for the end-user.
For the latter, the intervention developer could download the video in batches (stream in intervals), then apply automated video editing operations on each batch, and re-displaying it on the app; the operations would be undergoing while the app is streaming/loading the video. The latency to watch a single YouTube video would be noticeable in this instance. 
This processing method would need to be implemented for every video-based app, account for mobile browsers, as well as version change. 
App developers may also perform content integrity checks to safeguard their platform's content creator's content integrity. 

\begin{figure} 
    \centering
    \includegraphics[width=\linewidth]{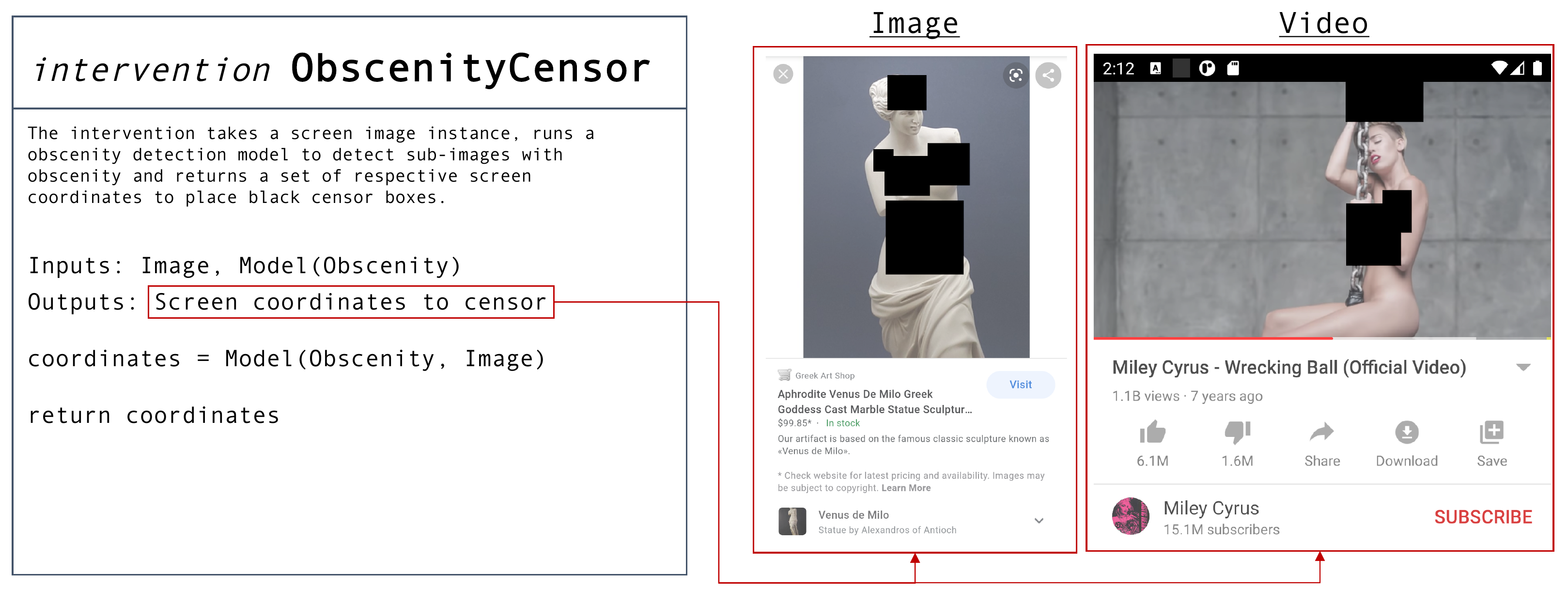}
    \caption{\gt{} can potentially help parents protect their children better, by filtering obscene content in apps in real time. The used ML model clearly leads to \textit{overblocking}, but this model can potentially be replaced with a more fine-grained one in the future. \gt{} aims to help researchers in finding better such models, with the help of end-users.}
    \label{fig:nud}
\end{figure}

\noindent{\small \textbf{Implementation with \gt{}: }}
An intervention developer can use the provided model hook, and combine this with an appropriate video analysis model (Figure~\ref{fig:nud}).
We chose a pre-trained nudity detection model~\citep{nn} to detect nudity in any screen image instance, and return a set of coordinates of each nude subimage.
A filter box can then be rendered over those coordinates to cover up.
Without \gt{}, an intervention developer may resort to processing incoming video streams, which can be cumbersome.
\gt{} provides a uniform approach to filter both images and videos in real-time, and across apps and platforms.

\section{Discussion}
\label{dis}


\subsection{Intervention developers: More time for research and design, less implementation}

Despite the existence of digital harms and other digital harms on mobile devices, it is challenging to deploy and test what appropriate intervention design should be.
Existing app modification frameworks require high implementation workload from the intervention developer, and make it difficult to scale intervention design experiments across a multitude of apps.
Even if an intervention design is proven to be successful, the intervention design can be difficult to scale, with considerations of robustness, such as change in app version. 
As shown in our five above case studies, we have greatly reduced the difficulty and time in implementing and deploying interventions. This means that intervention developers, designers and researchers can work on testing and evaluating the optimal interventions that should be deployed to users, not just porting interventions that are known to be effective on desktop browsers. They can spend more time on design and doing research, and less time on implementation.
Making interventions interface-centered, as done with \gt{}, allows them to be adaptive and employed dynamically across varying contexts to improve the overall interventions experience for the user.
Many prior app modification tools focused on app functionality, and thus needed to update interventions based on new app updates, new apps being released on the Google Play Store, porting mobile browser extensions to apps and vice versa. Now the intervention developer can use a single interface-oriented intervention to mitigate many harms swiftly, and managing and installing interventions have become simplified for the user.
Once an intervention has been developed, it can be relatively easily updated to fit different user needs, e.g. by changing the underlying ML model.
We show that we ease development of interventions by enabling scalability of an intervention implementation to different interfaces and needs.



One large opportunity is using machine learning as a tool to construct interventions, and also combating digital harms that use machine learning.
With the large accumulation of models on platforms such as Kaggle or ModelZoo, and open-sourcing of pre-trained models of deep learning research, there are plenty of models that can be translated to reality through \gt{}. 
Models trained to tackle semantic segmentation~\citep{5557884} could allow for granular object-specific filtering, and models trained to translate scenes to text~\citep{radford2021learning} can help identify scenes that may be likely to implicitly influence a person and may need warning labels in real-time.



\subsection{End-users: More autonomy over app interfaces?}








\gt{} takes a role in attempting to shift control of user interfaces from app developers to users.
To get access to the features of \gt{}, a user only needs to install a single app, rather than installing many separate solutions (with different installation and usage instructions) for each social media app or mobile browser.
There is no need to deal with the inconsistencies in performance between different interventions, as an intervention developer would only need to implement one intervention that can scale to different interfaces (rather than relying on separate interventions for different apps, with potentially varying quality).
\gt{} can be disabled at any time and without delay.
For example, when the stories bar is out of view, the output action taken (e.g. inpainting the stories bar) is stopped, and does not interfere with normal user operation unless the intervention is triggered. 
The usage of \gt{} requires no privilege escalation or root access; the end-user only needs to accept permissions to draw over other apps.
No app or system code needs to be modified; all app and device functionality is retained.



On the other hand, interventions remain subjective and personal.
What is the right balance between autonomy and paternalism \citep{Vandeveer2014} in regulating digital experiences?
Intervention tools, such as \gt{}, could be considered paternalistic to the end-user. 
If interventions have been developed by someone other than the end-user, 
the user has indirectly delegated autonomy over its device screen from app developers onto intervention programs and their developers. 
An example would be overblocking of content: some instances of media/text moderation could be considered adequate or overblocked depending on individual background and context.
This has been underlined by our case studies.
Another example is how broadly models are applicable: models trained on a single dataset may not be suit all mobile users. 
This is why, ideally, end-users should be able to develop their own interventions, without the need for third-party intervention developers, so as to retain full autonomy over their own device interfaces. 

Intervention tools are taking steps towards increased end-user autonomy, but they struggle to offer users full autonomy. 
This does not mean that we should abolish the category of interventions entirely. Rather, this underlines that the outsourcing of intervention regulation and digital autonomy needs high trust of the user in the developer regulating their interface.
While interventions could be argued to be paternalistic, the original app layout and interface could be considered as similarly -- if not more -- paternalistic, given that it may nudge users to exhibit certain behaviour without giving users the option to change their interface. 

Some interventions may need to be applied to the entire distribution of users if the benefit of paternalism outweighs the cost of autonomy, for example in cases of moderating suicidality or digital self-harm \citep{10.1145/3359186}; this same level of paternalism and justification applies to app developers who filter content on their platforms for all users in similar cases.
In cases where app developers have internal limitations or conflicting interests with users, users would need to find another source for digital regulation, and this is where interventions -- as described in this work -- step in. 

\subsection{Limitations}

Although we provide a framework of interface-centered interventions development, our current prototype faces a set of deployment limitations that future implementations may wish to address. 
Provided the user deploys a server on their local machine, 
the solution would not work when users are offline and would require high-bandwidth connectivity to support smooth operation. 
Continuously streaming the screen may affect battery life as well.

Intervention developers can make use of interventions and screen streaming to steal data or passwords from the end-user. 
This can be mitigated by having interventions distributed on a platform.
Users install a clean and verified app. 
If they install malicious interventions, the interventions can only introduce malicious graphics (which can be inspected and reported by other platform users). 
Intervention developers cannot use the graphics rendering interventions to perform on-device actions, nor do they have access to the screen images of the user. 
Additionally, 
end-users could run the framework on their local device, or the app could anonymize/censor screen content on the device before sending to a server, or the app could automatically turn off streaming and overlays in cases of sensitive content.

Another limitation pertains to the portability of \gt{} to iOS. For many popular apps, the versions on Android and iOS share overlapping interface elements (e.g. stories bar), so many masks and hooks should be compatible on iOS if the intervention tool \gt{} can be ported to iOS. Unfortunately this may not be the case in the near term due to a current lack of overlay capabilities for third-party apps on iOS.
However, if Apple opens up the overlay feature for developers (e.g. to facilitate accessibility tools), \gt{} can be easily ported to iOS.

\section{Conclusions}
\label{conclusions}

Digital harms are widespread, and affect individuals worldwide. Meanwhile, there exists limited available technology for researchers to understand what harms cause particular damage to individuals using mobile devices, and how to mitigate these harms effectively.
An important reason is the difficulty in developing and deploying interventions on mobile devices, partly due to the closed nature of the mobile digital ecosystem and the dominance of mobile software development by Apple and Google~--~each with their own philosophy and interests~\cite{kollnig2022_iphone_android}.

This work introduces a framework~--~which we call \gt{}~--~that allows researchers to develop, deploy, and test different interventions against interface-based digital harms easily, and thereby establish the ground truth around what interventions work for end-users.
\gt{} eases the development of interventions by providing a range of \textit{hooks}~--~ready-to-use templates~--~that can be used by researchers to build their own interventions.
We demonstrated 
the broad range of potential harms covered 
with a total of five cases studies of implemented interventions,
showing both adaptability and adoptability in the app modification framework.


\bibliographystyle{ACM-Reference-Format}
\bibliography{main}


\end{document}